\documentclass[twocolumn]{revtex4}
\usepackage[dvips]{graphicx}
\usepackage{float,color}

\begin{document}

\title{Stripe Order in Superfluid $^3$He Confined in Narrow Cylinders }

\author{Kazushi Aoyama ${}^{1,2}$}

\affiliation{${}^1$ The Hakubi Center for Advanced Research, Kyoto University, Kyoto 606-8501, Japan \\
${}^2$ Department of Physics, Kyoto University, Kyoto 606-8502, Japan 
}

\begin{abstract}
We theoretically investigate pairing states of the spin-triplet $p$-wave superfluid $^3$He confined in narrow cylinders. The surface-induced distortion and the multiple internal degrees of freedom of the order parameter lead to the occurrence of a stripe structure along the cylinder axis in the superfluid $^3$He-B phase. We show that in sufficiently small cylinders with an anisotropic surface scattering, the stripe order with broken translational symmetry may be stabilized as the lowest energy state. Periodic spin-current textures caused in this inhomogeneous superfluid phase are also discussed.  
\end{abstract}
\maketitle

The superfluid phases of liquid $^3$He are in spin-triplet $p$-wave Cooper pairing states. Because of the multiple spin and orbital degrees of freedom of Cooper pairs, superfluid $^3$He provides a rich variety of examples of symmetry breaking which is commonly seen in various contexts in condensed matter physics such as superconductors, magnets, and liquid crystals.  
The superfluid state associated with broken gauge symmetry is usually translationally invariant, as Cooper pairs condense into the coherent quantum state with total momentum ${\bf q}$ equal to zero, similarly to a Bose-Einstein condensation. 
In this paper, we address a special case with spontaneous translational symmetry breaking in superfluid $^3$He.
    
 
The bulk liquid $^3$He is a highly clean homogeneous system, and can potentially host various types of pairing states owning to the internal degrees of freedom. However, only two superfluid phases are realized in the bulk ${}^3$He, A and B phases which are respectively identified with the Anderson-Brinkman-Morel (ABM) and the Balian-Werthamer (BW) pairing states \cite{VW}.  
One possible way to stabilize different types of pairing states would be confinement of the superfluid in geometrically restricted spaces such as slabs, cylinders, and porous mediums such as aerogels. Scattering of $^3$He particles by surfaces of confining vessels and strands of the aerogels causes bending and pinning of the orbital components of the superfluid order parameter. In films of the $^3$He-B phase, the surface-induced distortion of the superfluid gap and a resultant transition into the A/planar phase have been suggested \cite{film_Nagato, film_Sauls, film_Murakawa, film_Science, film_PRL}. In aerogels, pinning effects in the ABM pairing state have been discussed \cite{Volovik_aero1, AI_is, AI_anis, Kunimatsu, Bunkov, Volovik_aero2}, and the polar pairing state which does not exist in the bulk is expected to appear in uniaxially stretched mediums \cite{AI_anis, Dmitriev}.
Also in narrow cylinders, occurrence of the polar state is theoretically predicted \cite{Fetter}, but so far, any signature of this state has not been observed in experiments \cite{cylinder_Manninen, cylinder_Kotsubo, cylinder_Pekola, cylinder_Saunders, cylinder_Yamaguchi}. 
Perhaps, the most intriguing question is whether or not there may be new pairing states other than BW, ABM, planar, and polar states in the restricted geometries. 
In this paper, we show that in the B phase confined in a sufficiently small cylinder, translational symmetry along the cylinder axis ($z$-axis) is spontaneously broken, i.e., $q_z$ becomes nonzero, as a result of the surface scattering.

From recent theoretical studies of the surface scattering effect, it is becoming clear that pairing states with spontaneous spatial modulations (${\bf q}\neq 0$) appear in thin films of unconventional superconductors \cite{Vorontsov_SC, Hachiya} and superfluid $^3$He \cite{Vorontsov_SF}. These spatially modulated phases are discussed by analogy to the Fulde-Ferrell-Larkin-Ovchinnikov (FFLO) state \cite{FF, LO} which was originally studied in the context of spin-singlet superconductors in a strong Zeeman field. 
The FFLO state is characterized by a nonzero ${\bf q}$ which is induced to compensate paramagnetic depairing, i.e., splitting of spin-up and down Fermi surfaces due to the Zeeman effect. 
In the film case without magnetic fields, bending of the order parameter near the film surface plays a role similar to the paramagnetic depairing, driving spatial modulations along the film plane. For the geometry of cylinder, the surface-induced distortion of the order parameter would trigger the FFLO-like instability. In superfluid $^3$He, however, the existence of the internal degrees of freedom is essential for the occurrence of the spatially modulated phase. 
As we will show below, in contrast to the FFLO state usually restricted to the quite low temperature region, spatially modulated stripe orders of superfluid ${}^3$He appear from relatively high temperatures below $T_c$, taking advantage of the internal degrees of freedom. This is our main finding of this work. 

Here we consider superfluid $^3$He in cylinders with radius $R$ less than the dipole length $\sim 12 \, \mu$m. For specular surface scattering, the ABM state is stable in a wide range of the $T-R$ phase diagram \cite{Li-Ho}, while for diffusive one, the BW state is favorable down to small $R \sim 5 \, \xi(T)$ [ $\xi(T)=\xi_0/(1-T/T_c)^{1/2}$ : temperature-dependent coherence length] followed by the second order transition into the polar phase at $T_{BP}(R)$ \cite{Fetter}. First, we will examine the FFLO-like instability in the BW state near $T_{BP}$. Next, relative stabilities of possible pairing states including the ABM state will be discussed for different two conditions of surface scatterings. Throughout this paper, we restrict ourselves to liquid $^3$He at 0 bar which is in the weak-coupling limit \cite{VW}. 
                                                                                               
Properties of liquid $^3$He near the superfluid transition temperature $T_c$ are well described by the Ginzburg-Landau (GL) theory \cite{VW}. As our focus is on pairing states in micron or submicron cylinders, we will neglect the dipole interaction as well as $\phi$-dependence of the order parameter.
The corresponding functional in the cylindrical coordinate system is obtained as an expansion in the spin-triplet $p$-wave order parameter $A_{\mu i}(r,z)$ \cite{Buchholtz_cylinder} 
\begin{eqnarray}\label{eq:GL}
{\cal F}_{\rm GL}&=&2\pi \int_{0}^{L_z}dz \int_0^R r \, dr \Big( f_{\rm bulk} + f_{\rm grad}\Big), \\
f_{\rm bulk} &=& \alpha A_{\mu i}^\ast A_{\mu i}+\beta_1 |A_{\mu i}A_{\mu i}|^2+\beta_2 (A_{\mu i}^{\ast}A_{\mu i})^2 \nonumber\\
&+& \beta_3 A_{\mu i}^{\ast}A_{\nu i}^{\ast}A_{\mu j}A_{\nu j} + \beta_4 A_{\mu i}^{\ast}A_{\nu i}A_{\nu j}^{\ast}A_{\mu j} \nonumber\\
&+& \beta_5 A_{\mu i}^{\ast}A_{\nu i}A_{\nu j}A_{\mu j}^{\ast} , \nonumber\\
f_{\rm grad}&=&  K_1 A^*_{\mu i,i}A_{\mu j,j} + K_2 A^*_{\mu i,j} A_{\mu i,j}+K_3 A^*_{\mu i,j}A_{\mu j,i} \nonumber\\
&+& 2r^{-1}{\rm Re}\big\{ K_1 (A^*_{r \phi}A_{\phi i,i}-A^*_{\phi \phi}A_{r i,i}+A^*_{\mu r}A_{\mu i,i}) \nonumber\\
&+& K_3 (A^*_{r i}A_{\phi \phi,i}-A^*_{\phi i}A_{r \phi,i}-A^*_{\mu \phi}A_{\mu \phi,r}) \big\} \nonumber\\
&+& r^{-2} \Big[ (K_1+K_3) ( |A_{r \phi}|^2 + |A_{\phi \phi}|^2 +A^*_{\mu r}A_{\mu r} ) \nonumber\\
&+& 2(K_1+K_3+2K_2){\rm Re}\big\{ A^*_{r \phi}A_{\phi r} - A^*_{r r}A_{\phi \phi}\big\}  \nonumber\\
&+& K_2( 2A^*_{\mu i}A_{\mu i} - A^*_{z i}A_{z i} - A^*_{\mu z}A_{\mu z} )\Big] , \nonumber
\end{eqnarray} 
with $A_{\mu i,j} \equiv \Big(\frac{\partial A_{\mu i}}{\partial r} , \, \frac{1}{r}\frac{\partial A_{\mu i}}{\partial \phi} , \, \frac{\partial A_{\mu i}}{\partial z} \Big)_j $.
In the weak coupling limit, the coefficients are given by $\alpha=\frac{1}{3}N(0)\ln(T/T_c)$, $-2\beta_1=\beta_2 = \beta_3 = \beta_4= -\beta_5 =2\beta_0$, $\beta_0\equiv 7\zeta(3)N(0)/(240\pi^2T^2)$, and $K_1=K_2=K_3=K\equiv \frac{1}{5} N(0) (T_c/T)^2 \xi_0^2$ with $N(0)$ as the density of state per spin on the Fermi surface and $\xi_0 = (v_{\rm F}/4 \pi T_c) \sqrt{7\zeta(3)/3}$ as the superfluid coherence length at $T=0$.
The presence of $r^{-2}$ centrifugal barriers in $f_{\rm grad}$ identifies the polar axis as a singular line, imposing constraints on the order parameter at $r=0$. This requires $A_{rr}=A_{\phi\phi}$ and $A_{rz}=A_{zr}=A_{z\phi}=0$ at $r=0$ for the BW and ABM states considered below.

The effect of the cylinder surface can be incorporated by the boundary condition on the order parameter $A_{\mu i}$. For specular surface scattering, the following condition must be satisfied at $r=R$ \cite{Buchholtz_cylinder}
\begin{equation}\label{eq:boundary_sp}
A_{\mu r}=0, \quad \frac{\partial A_{\mu \phi}}{\partial r}=\frac{1}{R}A_{\mu \phi}, \quad \frac{\partial A_{\mu z}}{\partial r}=0.
\end{equation}
This ensures that the normal component of the current vanishes at the walls. When the cylinder surface is sufficiently rough such that particles are randomly scattered independent of the incident direction, the boundary condition Eq.(\ref{eq:boundary_sp}) is replaced with the diffusive one, i.e., $A_{\mu i}|_{r=R}=0$ for any $\mu$ and $i$. 

\begin{figure}[t]
\includegraphics[scale=0.4]{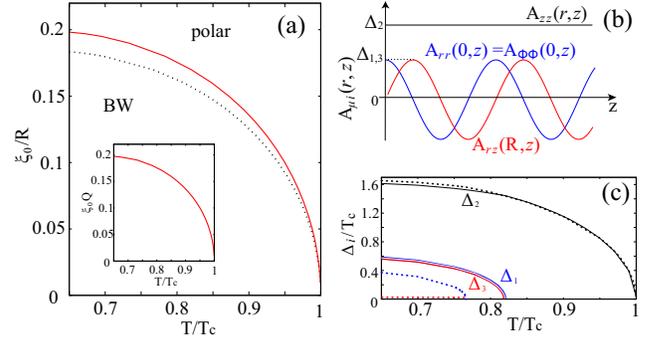}
\caption{(color online) (a) The BW-polar transition curves $T_{BP}(R)$ in cylinders with radius $R$ and the specular surface. $T_{BP}$ for the BW state with a spatial modulation along the cylinder axis (red solid curve) is higher than that for the uniform BW state (black dotted one). Inset shows the evolution of the modulation $Q$ along the $T_{BP}(R)$ curve. (b) Sketch of the spatial profile of the order parameter in the stripe BW state with the spatial modulation along $z$-axis. (c) Temperature dependences of $\Delta_1$ (blue), $\Delta_2$ (black), and $\Delta_3$ (red) at $\xi_0/R=0.17$ for the modulated (solid curves) and uniform (dotted ones) BW states. 
\label{fig:1}}
\end{figure}

Now, we turn to the stability of the BW state. In the BW state in the cylinder, the order parameter is usually assumed to be uniform along the cylinder axis ($z$-axis), and its basic form is given by $A^{(B)}_{\mu i}(r)=a_{rr}(r)\hat{r}_\mu\hat{r}_{i}+a_{\phi\phi}(r)\hat{\phi}_\mu\hat{\phi}_{i}+a_{zz}(r)\hat{z}_\mu\hat{z}_{i}$. It has been known that due to the surface-induced distortion of $a_{\mu i}(r)$, the BW state is gradually deformed into the polar state $A_{\mu i}(r)=a_{zz}(r) \hat{z}_{\mu}\hat{z}_{i}$ as cylinder radius $R$ shrinks \cite{Fetter, Li-Ho}. The basic assumption of this study is that the uniform BW state is described by $A^{(B)}_{\mu i}(r)$. To examine possibility of translational symmetry breaking in the $z$-direction, we take $z$-dependences of $A_{\mu i}$ into account, replacing the diagonal components $a_{\mu i}(r)$ with $z$-dependent ones $a_{\mu i}(r,z)$. With this replacement, however, FFLO-like states are not obtained at least within the GL theory. Instead, we notice that the $z$-dependent $A_{rr}$ and $A_{\phi \phi}$ couple with the additional component $A_{rz}$ through the gradient term $K{\rm Re}\{ A^\ast_{rz,r}A_{rr,z}+r^{-1}A^\ast_{rz}A_{\phi \phi,z} \}$. As we will see below, this gradient term connecting $A_{rz}$ with $A_{rr}$ and $A_{\phi \phi}$ gives rise to stripe ordering. We stress that this mechanism of the modulated state, i.e., coupling between the original components of the order parameter and the additional one, is specific to superfluid $^3$He with the internal degrees of freedom. In this study, based on the above discussion, we assume that the order parameter in the BW state takes the form of $A_{\mu i}(r,z)=A^{(B)}_{\mu i}(r,z)+a_{rz}(r,z)\hat{r}_\mu \hat{z}_i $ \cite{form}. When uniformity is assumed in the $z$-direction, the additional component $a_{rz}$ is found to vanish for boundary conditions used in this paper, suggesting that the existence of $a_{rz}$ leads to the energy costs for the uniform state. Hereafter, we will examine a single-mode instability and express the order parameter as   
\begin{eqnarray}\label{eq:BW_op}
&& A_{\mu i}(r,z) \\
&& = \left( \begin{array}{ccc}
a_{r r}(r)\cos(Qz) & 0 & a_{r z}(r)\sin(Qz) \\
0 & a_{\phi \phi}(r)\cos(Qz) & 0 \\
0 & 0 & a_{z z}(r) 
\end{array} \right), \nonumber
\end{eqnarray}
where $Q=2\pi n/L_z$ is the wave vector characterizing the spatial modulation of the order parameter along the cylinder axis ($z$-axis). Other combinations of the $z$-dependences of $A_{\mu i}$ including simple phase modulations $e^{\pm i Qz}$ do not yield a $Q$-linear gradient term, and thus, do not lead to modulated states  (see the text below). We also note that the cylindrical geometry requires the modulation to be along the $z$-axis. This is in contrast to the film case where possibility of complex spatial structures with two dimensional modulations cannot be ruled out. In what follows, we demonstrate that near the BW-polar transition $T_{BP}$, the BW state lowers the energy by introducing the modulation $Q$ and extends its stability region to higher temperatures. 

First, to get an insight into the mechanism of the modulated BW state, we consider the case with specular surface scattering and take the following trial state satisfying Eq.(\ref{eq:boundary_sp}),
\begin{eqnarray}\label{eq:OP_trial}
&& a_{r r}(r) = \Delta_1 \cos\big(\frac{\pi r}{2 R} \big), \quad a_{\phi \phi}(r) = \Delta_1 \frac{\cosh(\frac{r}{ R}) -e^{-1}}{1-e^{-1}} \nonumber\\
&& a_{z z}(r) = \Delta_2, \quad a_{r z}(r) = \Delta_3  \sin\big(\frac{\pi r}{2 R} \big) .
\end{eqnarray}
Translational symmetry breaking in the BW state is signaled by finite values of $Q$ and $\Delta_{1,3}$. In fact, nonzero values of $Q$ and $\Delta_{1,3}$ indicate {\it amplitude} modulations of the order parameter, similar to the LO state [see Fig.\ref{fig:1}(b)].
Inserting the expression Eq.(\ref{eq:OP_trial}) into Eq.(\ref{eq:GL}) and integrating over $r$ and $z$, we obtain
\begin{eqnarray}\label{eq:GL_trial}
{\cal F}_{\rm GL}&=& \pi R^2 L_z \Big[ \alpha \sum_i a_i \Delta_i^2 +\beta_0 \sum_{i\leq j} b_{ij} \Delta_i^2 \Delta_j^2 + K f_g  \Big], \nonumber\\
f_g &=& \Delta_1^2 \Big(\frac{g_1}{R^2}+g_1'Q^2 \Big) +\Delta_3^2 \Big( \frac{g_3}{ R^2} +g_3'Q^2 \Big) + g_{c} \Delta_1 \Delta_3 \frac{Q}{R} \nonumber\\
&=& D \Big( Q+\frac{g_{c}\Delta_1 \Delta_3}{2D R} \Big)^2 +\frac{\Delta_1^2}{R^2}\Big( g_1  + g_3 \frac{\Delta_3^2}{\Delta_1^2} -\frac{g_{c}^2 \Delta_3^2}{4D}\Big) 
\end{eqnarray}
with $D=g_1' \Delta_1^2 + g_3' \Delta_3^2$. The coefficients are calculated as $a_1=1.185$, $a_2=1$, $a_3=0.351$, $b_{11}=5.912$, $b_{12}=2.370$, $b_{13}=0.505$, $b_{22}=3$, $b_{23}=2.108$, $b_{33}=0.650$, $g_1=7.120$, $g_1'=1.185$, $g_3=1.191$, $g_3'=1.054$, and $g_c=-4.464$.
It should be emphasized that the gradient term linear in $Q$ shows up and consequently, the modulation takes the nonzero value $Q=-g_{c}\Delta_1 \Delta_3/(2D R)$ as long as $\Delta_3$ is nonzero to lower the gradient energy $f_g$ . This situation is in sharp contrast to conventional FFLO states for which a $Q$-linear gradient term does not exist and nonzero $Q$ appears because the coefficient of the $Q^2$ term becomes negative at very low temperatures for the strong depairing effect \cite{Adachi}. The present system is, however, rather similar to non-centrosymmetric superconductors (NCS) with Rashba spin-orbit coupling \cite{Rashba, Sigrist_book} in a magnetic field where the so-called helical phase with a field-induced phase modulation $e^{i {\bf Q}\cdot {\bf r}}$ is believed to be realized \cite{Dimitrova, Samokhin, Kaur, AS}. In Rashba-type NCS, broken inversion symmetry allows a $Q$-linear gradient term coupled with the field, and as a result, the phase modulation exists even at the superconducting transition temperature in the magnetic field. From the analogy to NCS, it is inferred that the modulated $^3$He-B phase would emerge from high temperatures near $T_c$.  

\begin{figure}[t]
\includegraphics[scale=0.6]{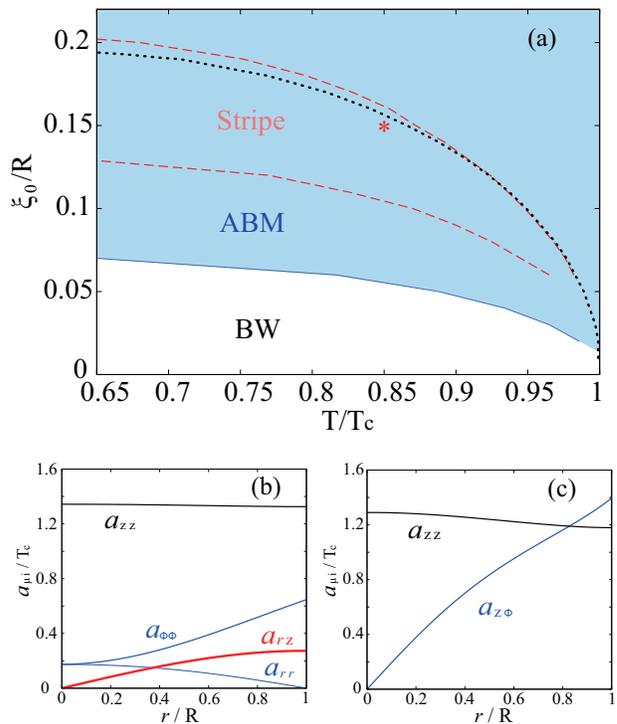}
\caption{(color online) Pairing states of superfluid $^3$He confined in narrow cylinders with radius $R$ and the specular surface. (a) T-R$^{-1}$ phase diagram obtained by numerically solving GL equations. The ABM state is realized in the blue colored area, while in the white colored region at large $R$'s, the BW state uniform along the cylinder axis is stable. For specular surface scattering, the stripe order with the axial modulation is not stable against the ABM state, and thus the lower and upper instability curves for the stripe phase (red dashed curves) as well as the $T_{BP}(R)$ curve for the uniform BW state (dotted one) are marginal. Radial dependences of the order parameters in the stripe phase and the ABM states at $\xi_0/R=0.15$ and $T/T_c=0.85$ (the symbol * in (a)) are shown in (b) and (c), respectively. \label{fig:2}}
\end{figure}

Figure\ref{fig:1} (a) shows the $T_{BP}(R)$ curves for the BW states of the form Eq.(\ref{eq:OP_trial}) with (red solid curve) and without (black dotted one) a modulation along the cylinder axis. The $T_{BP}$ transition temperature is enhanced for $Q \neq 0$, which implies that the superfluid state with broken translational symmetry is stabilized as the lowest energy BW state. Moreover, it is striking that it can survive up to relatively high temperatures. As seen in Fig.\ref{fig:1} (c), $\Delta_1$ and $\Delta_3$ grow up with the $\sqrt{|T-T_{BP}|}$ dependence suggestive of the second order BW-polar transition. Below $T_{BP}$, the comparable two components $\Delta_1 \simeq \Delta_3$ indicate $Q\simeq -g_c/[2(g_1'+g_3')R] \simeq 1/R$, which is in agreement with the obtained numerical result shown in the inset of Fig.\ref{fig:1} (a). For the optimized $Q$, we find that the gradient energy is lowered : $f_g -g_1\Delta_1^2/R^2=\{g_3-g_c^2/[4(g_1'+g_3')] \} \Delta_3^2/R^2 \simeq -\Delta_3^2/R^2$. Thus, the gradient term $f_g$ favors $\Delta_3$ together with $Q$. The result obtained by Eq.(\ref{eq:GL_trial}) is valid only when the trial state Eq.(\ref{eq:OP_trial}) well describes the exact spatial profile of the order parameter. For completeness, we numerically solve GL equations $\delta {\cal F}_{\rm GL}/\delta A^*_{\mu i}=0$ for the order parameter defined by Eq.(\ref{eq:BW_op}) at fixed points of $T$ and $R$, and determine the lowest energy state to obtain the phase diagram within the realm of the GL theory.    
Also, we take into account the ABM state of the form $A_{\mu i}(r)= \hat{z}_{\mu} \big[ \,  i\, a_{z r}(r)\hat{r}_{i}+ i\, a_{z \phi}(r)\hat{\phi}_{i} + a_{zz}(r) \hat{z}_{i} \big]$ \cite{Fetter, Li-Ho}. 

Figure \ref{fig:2} shows the numerically obtained whole phase diagram of superfluid ${}^3$He in cylinders with specular surface scattering. In the region sandwiched by red dashed curves, the stripe BW state with the modulation $Q\neq 0$ is more stable than the uniform one with $Q = 0$, where the lower red curve corresponds to the transition into the state with $Q=\pi/L_z$ \cite{Lower}. This stripe order in the BW state, however, is not stable against the ABM state. 
As shown in Fig.\ref{fig:2} (a), the $Q\neq 0$ region is completely masked by the ABM stability region, and the transitions into the modulated BW state are marginal for the purely specular surface scattering.

\begin{figure}[t]
\includegraphics[scale=0.6]{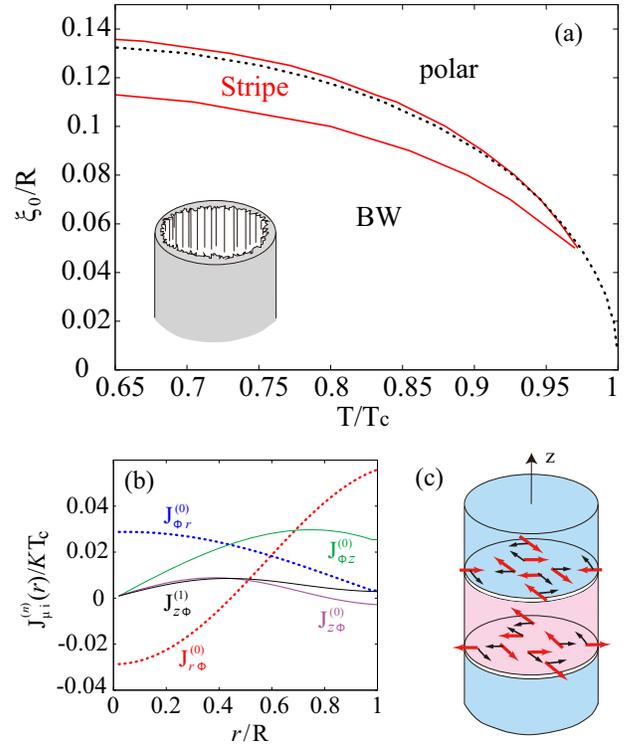}
\caption{(color online) Stripe phase realized in narrow cylinders with the anisotropic surface scattering described by Eq.(\ref{eq:boundary_anis}). (a) T-$R^{-1}$ phase diagram obtained by numerically solving GL equations. The stripe phase with the spatial modulation along the cylinder axis is stabilized in the region sandwiched by red solid curves. The inset shows an example of a cylinder with the surface condition, Eq.(\ref{eq:boundary_anis}). (b) The structure of spin currents in this inhomogeneous state at $\xi_0/R=0.12$ and $T/T_c=0.79$ for which $\xi_0 Q=0.11$. The spatial modulation gives rise to the periodic planar spin currents $J_{r\phi}$ and $J_{\phi r}$ as well as the oscillating behavior in $J_{\phi z}$ and $J_{z \phi}$. (c) Sketch of the planar spin currents flowing in opposite directions on adjacent nodal planes of $A_{rr} \, (A_{\phi\phi})$. Blue and pink colored regions denotes positive and negative values of $A_{rr} \, (A_{\phi\phi})$, respectively. Red (black) arrows represent the directions of the spin compontnt (flow). \label{fig:3}}
\end{figure}

The question is in which case the stripe phase may be stable as the lowest energy state. In realistic narrow cylinders, scattering at the wall is not perfectly specular because of roughness of the surface. Here, we consider a special case where the specularity of the surface has a directional anisotropy. As a simplified model for such an anisotropic surface scattering, we take the following boundary condition at $r=R$,
\begin{equation}\label{eq:boundary_anis}
A_{\mu r}=0, \quad A_{\mu \phi}=0, \quad \frac{\partial A_{\mu z}}{\partial r}=0.
\end{equation}
Namely, the surface of the cylinder vessel is assumed to be diffusive and specular along the $\phi$- and $z$- directions, respectively.  
An example of such an anisotropic cylinder wall is shown in the inset of Fig.\ref{fig:3} (a): the wall is smooth (rough) in the $z$- ($\phi$-) direction, and thus, at the wall particles are scattered in the forward direction (random directions) along the cylinder axis (rim). In the thin dirty layer model for surface roughness obtained by superposing a layer of randomly distributed impurities on a specular surface \cite{TDL1, TDL2}, the anisotropic surface structure would correspond to the case where the mean free path in the $z$-direction is much longer than that in the $\phi$-direction. In this situation, the $\phi$-components of the order parameter should be strongly suppressed at the wall while the $z$-components are almost unaffected, leading to a boundary condition similar to Eq.(\ref{eq:boundary_anis}). 
Even for diffusive cylinder walls, since their specularity can be tuned by coating their surfaces with $^4$He \cite{specularity}, such a directional anisotropy of the surface scattering would exist in the case that nonuniformity of the $^4$He coat is much larger in the $\phi$-direction. 
Since for the specular condition, $a_{z \phi}(R)$ in the ABM state is larger than $a_{\phi \phi}(R)$ in the BW state (see Figs.\ref{fig:2} (b) and (c)), suppression of both $a_{z\phi}$ and $a_{\phi\phi}$ due to the anisotropic surface scattering should lead to more significant reduction of the superfluid condensation energy in the ABM state than in the BW state.

Figure \ref{fig:3} shows the phase diagram of superfluid $^3$He in cylinders with the anisotropic surface scattering described by Eq.(\ref{eq:boundary_anis}). The ABM state becomes less stable for such an anisotropic scattering, and the stability region of the stripe BW state is unveiled. As one can see in Fig.\ref{fig:3} (a), the inhomogeneous stripe phase appears from relatively high temperatures in narrow cylinders with radius $R$ less than $20 \, \xi_0$.  
When the cylinder surface is diffusive in any direction, the stripe phase is not stable any more. Thus, the stability region of the stripe phase would be gradually suppressed as the surface scattering along the $z$-direction becomes more diffusive.  

Finally, we address physical properties of the stripe phase. Since the order parameter of this new phase takes real values, it has neither spontaneous current nor magnetization. Instead, the stripe structure yields spin currents periodic along the cylinder axis. The flux of spin component $\sigma$ in the direction $i$ has been derived elsewhere \cite{Leggett,jspin_Fomin,Buchholtz_cylinder} and is given for the BW state as
\begin{eqnarray}
J_{\sigma i} &=& -2K\epsilon_{\sigma\mu\nu}\Big( A_{\mu i}A_{\nu j,j}+ A_{\mu j} A_{\nu i,j}+A_{\mu j}A_{\nu j,i} \nonumber\\
&+& r^{-1}\big[A_{\mu i}A_{\nu r} -\delta_{\nu r}(A_{\mu i}A_{\phi\phi}+A_{\mu \phi}A_{\phi i}) \nonumber\\
&+& \delta_{i\phi}(2A_{\mu \phi}A_{\nu r}-A_{\mu r}A_{\nu \phi}-\delta_{\nu r}A_{\mu j}A_{\phi j} \nonumber\\
&+& \delta_{\nu \phi}A_{\mu j}A_{r j}) \, \big]\Big). 
\end{eqnarray}
In the stripe phase with the spatially modulation, there are four nonvanishing components, $J_{r\phi}=J^{(0)}_{r\phi}(r)\sin(Qz)$, $J_{\phi r}=J^{(0)}_{\phi r}(r)\sin(Qz)$, $J_{\phi z}=J^{(0)}_{\phi z}(r)\cos(Qz)$, and $J_{z\phi}=J^{(0)}_{z \phi}(r)+ J^{(1)}_{z \phi}(r)\cos(2Qz)$. Note that in the BW state uniform along the $z$-axis, we have only the two components $J_{\phi z}$ and $J_{z \phi}$. Radial dependences of the amplitude of these spin currents are shown in Fig.\ref{fig:3} (b). In the stripe phase, the additional components $J^{(0)}_{r\phi}$ and $J^{(0)}_{\phi r}$ are nonzero at $r=0$ suggesting the existence of the planar components $-J_{xy}=J_{yx}=J^{(0)}_{\phi r}(r=0)=-J^{(0)}_{r \phi}(r=0)$. As shown in Fig.3 (c), the planar spin current periodically changes its sign along the cylinder axis, and the longitudinal flow $J_{\phi z}$ exhibits an oscillating behavior as well. Such periodic textures of spin currents originating from the spatial variation of the $d$-vector should be in principle reflected in NMR frequency shifts.

To conclude, we find that a spatially inhomogeneous stripe order can emerge in superfluid $^3$He confined in a very narrow cylinder, due to both the surface-induced bending and the multiple internal degrees of freedom of the order parameter. Interestingly, this stripe order appears from high temperatures near $T_c$, and exhibits periodic spin-current textures along the cylinder axis whose signatures could be potentially observed in NMR experiments. 

The author is grateful to R. Ikeda for illuminating discussions and Y. Sasaki and Y. Lee for their useful comments. This work is supported by a Grant-in-Aid for Scientific Research (Grant No. 25800194).


\begin{thebibliography}{60}
\bibitem{form} Although other components of $A_{\mu i}$ may help the occurrence of the BW state with a spatial variation in the $z$-direction, the full calculation incorporating all the components is out of scope of our study.
\bibitem{Lower} The lower boundary of the stripe phase is determined by the full numerical calculation under the condition $A_{r r}(r,L_z)=-A_{rr}(r,0)$, $A_{\phi \phi}(r,L_z)=-A_{\phi \phi}(r,0)$, $A_{zz}(r,L_z)=A_{zz}(r,0)$, and $A_{rz}(r,L_z)=A_{rz}(r,0)=0$, without assuming Eq.(\ref{eq:BW_op}). The obtained spatial structure of the order parameter corresponds to that with $Q=\pi/L_z$ in Eq.(\ref{eq:BW_op}).
\end{thebibliography}
\end{document}